\documentclass[12pt,english,aps,pra]{revtex4}
\usepackage[T1]{fontenc}
\usepackage[latin9]{inputenc}
\usepackage{units}
\usepackage{amsmath}
\usepackage{graphicx}
\usepackage{amssymb}

\makeatletter
\@ifundefined{textcolor}{}
{%
 \definecolor{BLACK}{gray}{0}
 \definecolor{WHITE}{gray}{1}
 \definecolor{RED}{rgb}{1,0,0}
 \definecolor{GREEN}{rgb}{0,1,0}
 \definecolor{BLUE}{rgb}{0,0,1}
 \definecolor{CYAN}{cmyk}{1,0,0,0}
 \definecolor{MAGENTA}{cmyk}{0,1,0,0}
 \definecolor{YELLOW}{cmyk}{0,0,1,0}
 }

\usepackage{mathrsfs}\usepackage{bbm}
\makeatother

\usepackage{babel}
\usepackage{fancyhdr}

\makeatother

\usepackage{babel}

\begin{document}

\title{Modified-scaled hierarchical equation of motion approach for the
study of quantum coherence in photosynthetic complexes }

\author{Jing Zhu and Sabre Kais}

\thanks{Corresponding author, kais@purdue.edu}

\address{Department of Chemistry and Birck Nanotechnology Center, Purdue University,
West Lafayette, IN 47907, USA}

\author{Patrick Rebentrost and Al\'{a}n Aspuru-Guzik}

\address{Department of Chemistry and Chemical Biology, Harvard University,
12 Oxford Street, Cambridge, MA 02138, USA}
\begin{abstract}
We present a detailed theoretical study of the transfer of electronic
excitation energy through the Fenna-Matthews-Olson (FMO) pigment-protein
complex, using the new developed modified scaled hierarchical approach
{[}Shi Q. et al, J Chem Phys 2009, 130, 084105{]}. We show that this
approach is computationally more efficient than the original hierarchical
approach. The modified approach reduces the truncation levels of the
auxiliary density operators and the correlation function. We provide
a systematic study of how the number of auxiliary density operators
and the higher-order correlation functions affect the exciton dynamics.
The time scales of the coherent beating are consistent with experimental
observations. Furthermore, our theoretical results exhibit population
beating at physiological temperature. Additionally, the method does
not require a low-temperature correction to obtain the correct thermal
equilibrium at long times. 
\end{abstract}
\maketitle

\section{Introduction}

In the initial step of photosynthesis, light is captured by protein-bound
pigments that are part of light-harvesting antenna complexes. The
excitation energy is transferred with a near-unity quantum yield to
reaction centers, where it is converted to chemical energy. The underlying
molecular mechanisms responsible for the near-unity quantum yield
are far from being understood. In this regard, an extensively studied
and relevant system is the Fenna-Matthews-Olson (FMO) pigment-protein
complex of green-sulphur-bacteria, which acts as a mediator of excitation
energy between the outer antenna system, i.e., the chlorosomes, and
the reaction center \cite{Andrew-book}.

Experimentally, Savikhin et al. \cite{Savikhin-1997} observed quantum
beating in the FMO complex using the fluorescence anisotropy technique.
More recently, Engel et al. \cite{Engel-2007} employed two-dimensional
electronic spectroscopy to observe long-lasting quantum beats that
provides direct evidence for survival of long-lived electronic coherence
for hundreds of femtoseconds. Aspuru-Guzik et al. \cite{Alan-2008,Alan-2009,Alan-JPC-2009}
investigated the effects of quantum coherence and the fluctuating
environment, using the Lindblad formalism, on the enhancement of the
photosynthetic electronic energy transfer efficiency from the perspective
of a quantum walk. In Rebentrost et al. \cite{Alan-JPC-2009}, a method
to quantify the role of quantum coherence was introduced. Ishizaki
et al. \cite{Fleming-PNAS-2009,Fleming-JCP-2009,Fleming-JPC-2005}
employed the hierarchical equation of motion (HEOM) approach expansion
in order to address the robustness and the role of the quantum coherence
under physiological conditions. Their results reveal that quantum
wave-like motion persists for several hundred femtoseconds even at
physiological temperature $T=300K$. Very recently, a very large-scale
calculation of energy transfer between chromophore rings of purple
bacteria was carried out using the HEOM approach \cite{Strumpfer2009}.
Meanwhile, Whaley et al. \cite{Whaley-2009} and Caruso et al. \cite{Entanglement02}
discussed quantum entanglement in photosynthetic harvesting complexes
and clarified the connection between coherence and entanglement. They
showed that the FMO complex exhibits bipartite entanglement between
dimerized chromophores. The subject continues to be of great interest
with a large number of publications discussing electronic energy transfer
in photosynthetic complexes, in particular, the issue of quantum speed
up in the FMO complex \cite{Fleming-Science-2007,Sension-2007,Uri-2008,Cheng-Fleming-2009,Collini-2010,Scholes-2010,whaley}.

In this paper, we present a detailed theoretical study of the transfer
of excitation energy towards the reaction center through the Fenna-Matthews-Olson
(FMO) pigment-protein complex using a modified scaled hierarchy equation
of motion approach, which was developed by Shi and coworkers recently
\cite{Shi2009a}. This approach guarantees that the auxiliary density
operators decay to zero at high truncation level. Furthermore, it
provides a considerable computational speedup over the original hierarchical
approach \cite{Fleming-PNAS-2009}. We will show that the scaled hierarchical
approach can reduce the truncation level of both auxiliary density
operators and the correlation functions compared to the classical
approach. The time scales of the coherent beating are consistent with
experimental observations. Furthermore, our results show that the
population beating persists at physiological temperature.

\section{Theory}

Excitonic energy transfer within photosynthetic proteins -such as
the FMO complex- operates in a demanding parameter regime where a
small perturbative quantity is not available. It is thus a challenge
to find accurate and efficient methods for the simulation of the quantum
dynamics. A number of approximate methods have been developed \cite{Cheng-Fleming-2009}:
they include the semi-classical F\"{o}rster theory, standard Redfield
theory, modified Redfield theories, the modified Lindblad formalism
and hierarchical equations of motion, among others \cite{Tanimura-Kubo-1989,Tanimura-Wolynes-1991,Alan-2008}.

In this study, we utilize the recent hierarchical Liouville space
propagator method developed by Shi et al. \cite{Shi2009a} to investigate
excitation energy transfer in the FMO complex. The original method
is based on a reformulation of the original hierarchical quantum master
equation and the incorporation of a filtering algorithm that automatically
truncates the hierarchy with a preselected tolerance. They showed
how this method significantly reduces the number of auxiliary density
operators used to calculate electron transfer dynamics in a spin-boson
model and the absorption spectra of an excitonic dimer \cite{Shi2009a}.

\begin{figure}
\begin{centering}
\includegraphics[width=1\textwidth]{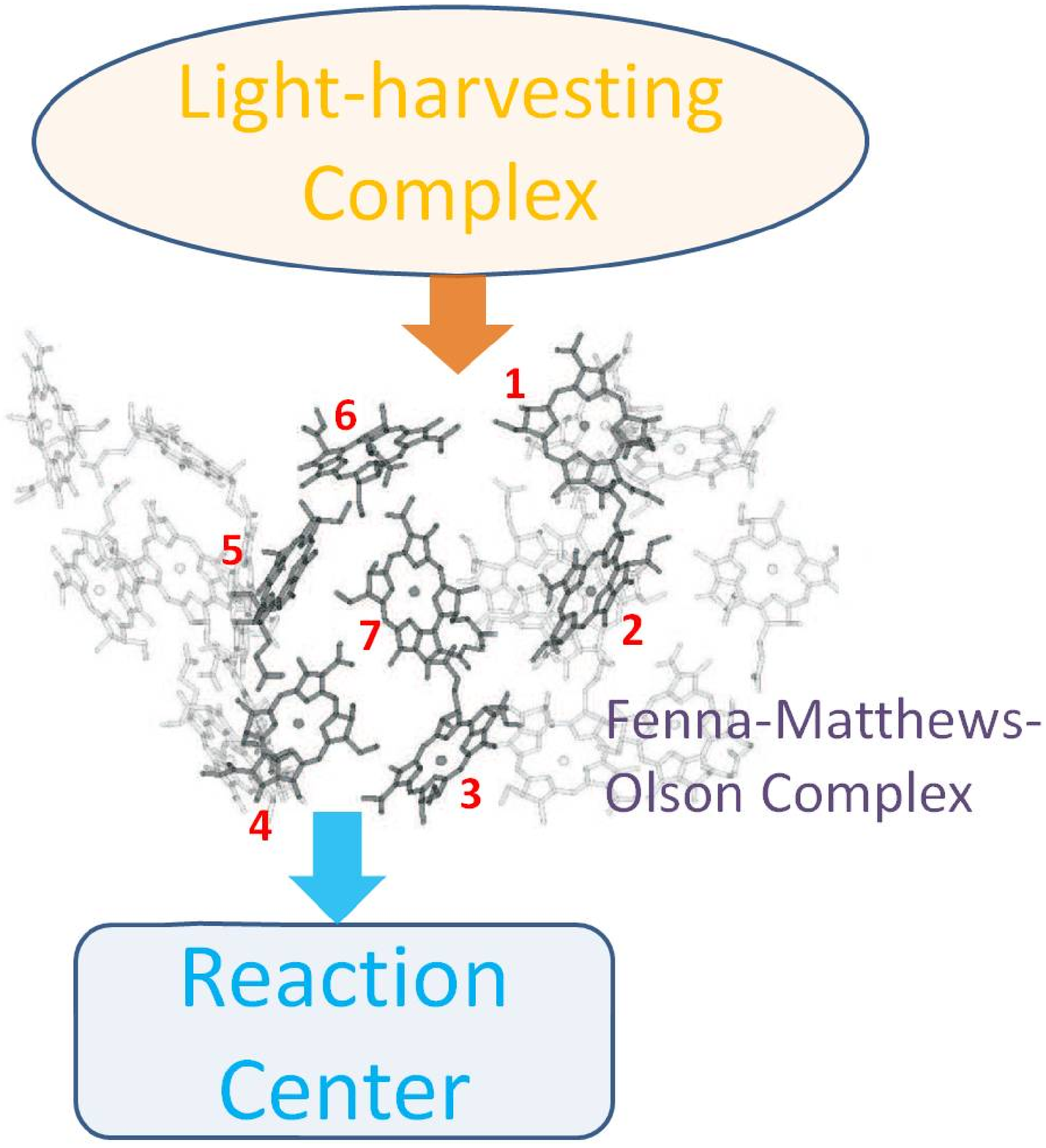} 
\par\end{centering}

\caption{Sketch of the energy flow in the process of photosynthesis. The energy
is captured by the light-harvesting complexes (LHC) and transferred
to the reaction center (RC). The FMO complex is the link between LHC
and RC and it operates as a \textquotedbl{}wire\textquotedbl{}
during the energy transfer process. Using the convention for numbering
the BChl a molecules (sites) of FMO complex as in ref. \cite{FENNA1975}
, site $1$ and $6$ are close to the LHC and site $3$ and $4$ are
close to the RC. In our theoretical description, the excited states
of the BChl a molecules of the FMO complex are considered as the \textquotedbl{}system\textquotedbl{}
and all the other relevant degrees of freedom are referred to as the
\textquotedbl{}environment\textquotedbl{}. \label{framework}}

\end{figure}

The structure of FMO complex was first analyzed by Fenna and Matthews
in 1975 \citep{FENNA1975}. It consists of a trimer, formed by three
identical monomers. Each monomer contains seven bacteriochlorophyll
a (BChl a) molecules (or seven {}``sites''.) Biologically, the FMO
complex acts as a molecular energy wire, transferring excitation energy
from the chlorosome structure, where light is captured, to the reaction
center (RC). There is substantial evidence that the FMO complex is
oriented such that sites $1$ and $6$ are close to the baseplate
protein and sites 3 and 4 are close to the RC complex and thus define
the target region for the exciton \citep{FENNA1975,Li1997,Camara-Artigas2003,Cheng-Fleming-2009}.
The detailed structure is shown in Fig. \ref{framework}. It should
be mentioned that the presence of the eighth BChl a molecules per
monomer has been proposed by Ben-Shem et al in 2004 \cite{Ben-Shem2004}.
This has been verified experimentally recently by Tronrud and coworkers
\cite{Tronrud2009}. It has been suggested that the eighth BChl a
molecule acts as a gateway site from the reaction center to the $27$
chromophores in the trimer \cite{Tronrud2009}.

The total Hamiltonian of the quantum system is given by:

\begin{equation}
\mathcal{H}=\mathcal{H_{S}}+\mathcal{H}_{B}+\mathcal{H}_{SB},\label{eq:htot}\end{equation}
 where $\mathcal{H}_{S}$, $\mathcal{H}_{B}$, and $\mathcal{H}_{SB}$
are the Hamiltonian of the system, the environment, and the system-environment
coupling, respectively. Here, we consider each site as a two-level
system of ground state and excited state. The system Hamiltonian,
$\mathcal{H}_{S}$, which describes the electronic states of the pigments
can be expressed as:

\begin{equation}
\mathcal{H}_{S}=\sum_{j=1}^{N}\varepsilon_{j}\,|j\rangle\langle j|+\sum_{j<k}J_{jk}\left(\,|j\rangle\langle k|+\,|k\rangle\langle j|\right),\label{eq:hs}\end{equation}
 where $|j\rangle$ denotes the state with only the $j$-th site is
in its excited state and all other sites are in their ground state.
$\varepsilon_{j}$ represents the site energy of the $j$-th site
which is defined as the optical transition energy at the equilibrium
configuration of environmental phonons associated with the ground
state. $N$ is the number of pigments or sites. $J_{jk}$ is the electronic
coupling between site $j$ and $k$. The parameters for this Hamiltonian
are taken from the paper of Adolphs and Renger \cite{Adolphs2006}.
In their work, two independent methods were used to obtain the site
energies of the seven BChl a molecules of the monomeric subunits of
the FMO complex. In the first method, the site energies are used as
parameters that were optimized by a genetic algorithm in the fit of
the optical spectra. In the second one, the site energies are obtained
directly by electrochromic shift calculations \cite{Adolphs2006}.

For the Hamiltonian of the environment, $\mathcal{H}_{B}$, a harmonic
oscillator model is applied. Furthermore, it is assumed that the electronic
excitation on each site couples to its own bath independently:

\begin{equation}
\mathcal{H}_{B}=\sum_{j=1}^{N}\mathcal{H}_{B}^{j}=\sum_{j=1}^{N}\sum_{\xi=1}^{N_{jB}}\frac{P_{j\xi}^{2}}{2m_{j\xi}}+\frac{1}{2}m_{j\xi}\omega_{j\xi}^{2}x_{j\xi}^{2},\label{eq:hb}\end{equation}
 where $N_{jB}$ is the number of different harmonic modes coupled
to the $j$-th site. Here, $m_{j\xi}$, $\omega_{j\xi}$, $P_{j\xi}$
and $x_{j\xi}$ are mass, frequency, momentum, and position operator
of the harmonic bath modes. The coupling term, $\mathcal{H}_{SB}$,
which is responsible for fluctuations in the site energies by the
phonon dynamics, can be expressed as:

\begin{equation}
\mathcal{H}_{SB}=\sum_{j=1}^{N}\mathcal{H}_{SB}^{j}=-\sum_{j=1}^{N}|j\rangle\langle j|\cdot F_{j}=-\sum_{j=1}^{N}\mathcal{V}_{j}\cdot F_{j},\label{eq:hsb}\end{equation}
 where the bath part is $F_{j}=\sum_{\xi}c_{j\xi}\cdot x_{j\xi}$
and the $c_{j\xi}$ represent the system-environment coupling constant
for the $j$-th site and $\xi$-th phonon mode. The projection operator
$\mathcal{V}_{j}=|j\rangle\langle j|$ describes the system part of
the interaction.

At time $t=0$, we assume that the system and the environment are
decoupled, i.e. $\rho_{tot}\left(0\right)=\rho\left(0\right)\otimes\rho_{B}\left(0\right)$.
Additionally, the environment is in the Boltzmann equilibrium state,
$\rho_{B}\left(0\right)=\nicefrac{\mathbbm{e}^{-\beta H_{B}}}{Tr_{B}\left[\mathbbm{e}^{-\beta H_{B}}\right]}$,
where $\beta=\nicefrac{1}{k_{B}T}$. The time evolution of the system
density matrix, $\rho\left(t\right)$, can be calculated by tracing
out the environment degrees of freedom:

\begin{equation}
\rho\left(t\right)=Tr_{B}\left[\rho_{tot}\left(t\right)\right]=Tr_{B}\left[\mathbbm{e}^{\nicefrac{-i\mathcal{H}t}{\hbar}}\,\rho_{tot}\left(0\right)\,\mathbbm{e}^{\nicefrac{i\mathcal{H}t}{\hbar}}\right].\label{eq:rho1}\end{equation}
 The bath is described by its correlation functions, $C_{j}\left(t\right)$,
which are defined as \citep{Ishizaki2005,Xu2005,Strumpfer2009}, $C_{j}\left(t\right)=Tr_{B}[\tilde{F_{j}}\left(t\right)\,\tilde{F_{j}}\left(0\right)\,\rho_{B}]$,
where the Langevin force, $\tilde{F_{j}}\left(t\right)$, is given
in the interaction picture: $\tilde{F_{j}}\left(t\right)=\mathbbm{e}^{\nicefrac{i\mathcal{H}_{B}t}{\hbar}}\, F_{j}\,\mathbbm{e}^{\nicefrac{-i\mathcal{H}_{B}t}{\hbar}}$.
For the phonon bath, the correlation function can be written as $C_{j}\left(t\right)=\frac{1}{\pi}\intop_{-\infty}^{\infty}d\omega\cdot J_{j}\left(\omega\right)\cdot\frac{\mathbbm{e}^{-i\omega t}}{1-\mathbbm{e}^{-\beta\hbar\omega}}$,
where $J_{j}\left(\omega\right)$ is the spectral density for the
$j$-th site:

\begin{equation}
J_{j}\left(\omega\right)=\sum_{\xi}\frac{c_{j\xi}^{2}\cdot\hbar}{2m_{j\xi}\cdot\omega_{j\xi}}\delta\left(\omega-\omega_{j\xi}\right).\label{eq: original spec}\end{equation}
 To proceed, we use the Drude spectral density, which corresponds
to an overdamped Brownian oscillator model. Furthermore, we assume
that the system-environment coupling is the same for all sites, $J_{j}\left(\omega\right)=J\left(\omega\right)$,
$\forall$ $j$s. The Drude spectral density is defined as:

\begin{equation}
J\left(\omega\right)=\eta\gamma\frac{\omega}{\omega^{2}+\gamma^{2}}.\label{eq:drude}\end{equation}
 We introduced $\eta=\frac{2\lambda}{\hbar}$ -which is dependent
on the reorganization energy $\lambda$- and the Drude decay constant,
$\gamma$. Under this spectral density, the correlation function $C_{j}\left(t\right)$
takes the form:

\begin{equation}
C_{j}\left(t>0\right)=\sum_{k=0}^{\infty}c_{k}\cdot\mathbbm{e}^{-v_{k}t},\label{eq:temperature extension}\end{equation}
 with the Matsuraba frequencies $v_{0}=\gamma$ and $v_{k}=\frac{2k\pi}{\beta\hbar}$
for $k\geqslant1$. The constants $c_{k}$ are given by:

\begin{eqnarray*}
c_{0} & = & \frac{\eta\gamma}{2}\left[cot\left(\frac{\beta\hbar\gamma}{2}\right)-i\right],\\
c_{k} & = & \frac{2\eta\gamma}{\beta\hbar}\cdot\frac{v_{k}}{v_{k}^{2}-\gamma^{2}}\;\; for\,\, k\geqslant1.\end{eqnarray*}
 Now, we are in the position to write down the HEOM for the reduced
density operator \cite{Strumpfer2009},

\begin{eqnarray}
\frac{d}{dt}\rho_{\boldsymbol{n}} & = & -\left(i\mathcal{L}_{S}+\sum_{j=1}^{N}\sum_{k}n_{jk}v_{k}\right)\rho_{\boldsymbol{n}}\label{eq: HEOM}\\
 &  & -i\sum_{j=1}^{N}\left[\mathcal{V}_{j},\;\sum_{k}\rho_{\boldsymbol{n_{jk}^{+}}}\right]\nonumber \\
 &  & -i\sum_{j=1}^{N}\sum_{k}n_{jk}\left(c_{k}\,\mathcal{V}_{j}\rho_{\boldsymbol{n_{jk}^{-}}}-c_{k}^{*}\,\rho_{\boldsymbol{n_{jk}^{-}}}\mathcal{V}_{j}\right),\nonumber \end{eqnarray}
 where $\boldsymbol{n}$ denotes the set of nonnegative integers $\boldsymbol{n}\equiv\{n_{1},n_{2},\cdots,n_{N}\}=\{\{n_{10},n_{11},\cdots,\, n_{1K}\},\cdots,\{n_{N0},n_{N1},\cdots,\, n_{NK}\}\}$.
$\boldsymbol{n_{jk}^{\pm}}$ refers to the change of the number $n_{jk}$
to $n_{jk}\pm1$ in the global index $\boldsymbol{n}$. The sum of
$n_{jk}$ is called tier ($\mathcal{N}_{c}$), $\mathcal{N}_{c}=\sum_{j,k}n_{jk}$.
In particular, $\rho_{\boldsymbol{0}}=\rho_{\{\{0,0,\cdots\},\cdots,\{0,0,\cdots\}\}}$
is the system's reduced density operator (RDO) and all others are
auxiliary density operators (ADOs). Although the RDO is the most important
operator, the ADOs contain corrections to the system-environment interaction;
these arised from the non-equilibrium treatment of the bath.

Here, we assume that both $\rho_{\boldsymbol{0}}$ and $\mathcal{V}_{j}$
have the order of one. When the tier of $\rho_{\boldsymbol{n}}$ at
tier $\mathcal{N}_{c}$ ($\mathcal{N}_{c}=\sum_{j,\, k}n_{jk}$ ),
the amplitude of $\rho_{\boldsymbol{n}}$ is proportional to $\left|c_{0}^{\sum_{j}n_{j0}}c_{1}^{\sum_{j}n_{j1}}\cdots c_{K}^{\sum_{j}n_{jK}}\right|$
following the standard approach (Eq. \ref{eq: HEOM}) \cite{Tanimura,Tanimura2},
which indicates the amplitude of $\rho_{\boldsymbol{n}}$ is related
to both $c_{k}$ and $n_{jk}$. The $n_{jk}$ is decided by the truncation
level, while the $c_{k}$ is related to the correlation function.
The correlation function is derived from the system-environment correlation.
In other words, the amplitude of $\rho_{\boldsymbol{n}}$ is dependent
on the system-environment coupling. Under the intermediate-to-strong
system-environment coupling, the amplitude of $\rho_{\boldsymbol{n}}$
can not be guaranteed to be small even at high truncation level. It
goes to the opposite direction as we expected as we always expect
more accurate results at high truncation level. Fortunately, Shi and
coworkers developed a new approach in which one is able to rescale
the original ADOs which can be used for overcoming this issue \cite{Shi2009a}.
They scaled the original operator as:

\begin{equation}
\tilde{\rho}_{\boldsymbol{n}}\left(t\right)=\left(\prod_{k,\, j}n_{jk}!\,\left|c_{k}\right|^{n_{jk}}\right)^{-\nicefrac{1}{2}}\rho_{\boldsymbol{n}}\left(t\right),\label{eq:scaled rho}\end{equation}
 After the scaling, the $\left|\tilde{\rho}_{\boldsymbol{n}}\right|$
has the order of $\prod_{k,\, j}\sqrt{\nicefrac{\left|c_{k}\right|^{n_{jk}}}{n_{jk}!}}$
. It can make sure that $\left|\tilde{\rho}_{\boldsymbol{n}}\right|$
decays to zero at higher hierarchical truncation level.

Since the number of contributing terms to the correlation function,
Eq. \ref{eq:temperature extension}, and ADOs are infinite, the computation
of Eq. \ref{eq: HEOM} is -in general- impossible. In order to overcome
this problem, a truncation scheme for both the correlation function
and ADOs is applied. We set the truncation level for the correlation
function (Matsuraba frequency and constant $c_{k}$) at level $K$,
while the cutoff for the tier of ADOs is $\mathcal{N}_{c}$. With
the Ishizaki-Tanimura truncating scheme \cite{Tanimura,Tanimura2},
Eq. \ref{eq: HEOM} for the scaled density operator becomes:

\begin{eqnarray}
\frac{d}{dt}\tilde{\rho}_{\boldsymbol{n}} & = & -\left(i\mathcal{L}_{S}+\sum_{j=1}^{N}\sum_{k=0}^{K}n_{jk}v_{k}\right)\tilde{\rho}_{\boldsymbol{n}}\label{eq:final HEOM}\\
 &  & -i\sum_{j=1}^{N}\sum_{k}\sqrt{\left(n_{jk}+1\right)\left|c_{k}\right|}\,\left[\mathcal{V}_{j},\;\tilde{\rho}_{\boldsymbol{n_{jk}^{+}}}\right]\nonumber \\
 &  & -\sum_{j=1}^{N}\sum_{m=K+1}^{\infty}\frac{c_{jm}}{v_{jm}}\cdot\left[\mathcal{V}_{j},\,\left[\mathcal{V}_{j},\,\tilde{\rho}_{\boldsymbol{n}}\right]\right]\nonumber \\
 &  & -i\sum_{j=1}^{N}\sum_{k=0}^{K}\sqrt{\nicefrac{n_{jk}}{\left|c_{k}\right|}}\;\left(c_{k}\mathcal{V}_{j}\,\tilde{\rho}_{\boldsymbol{n_{jk}^{-}}}-c_{k}^{*}\tilde{\rho}_{\boldsymbol{n_{jk}^{-}}}\mathcal{V}_{j}\right).\nonumber \end{eqnarray}
 We use Eq. \ref{eq:final HEOM} to simulate the exciton dynamics
of the FMO complex.

\section{Results and discussion}

For the numerical analysis, we used the same Hamiltonian as in Refs.
\cite{Fleming-PNAS-2009,Adolphs2006}, the same reorganization energy,
$\lambda_{j}=\lambda=35\;\unit{cm^{-1}}$, and Drude decay constant,
$\gamma_{j}^{-1}=\gamma^{-1}=50\;\unit{fs}$, of Ref. \cite{Read2008,Fleming-PNAS-2009},
As we mentioned before, sites $1$ and $6$ are both connected to
the LHC. It is possible that sites $1$, $6$ or both are excited.
For this reason, three different initial conditions are employed,
$|1\rangle$(site 1 is excited), $|6\rangle$ (site $6$ is excited)
and $\frac{1}{\sqrt{2}}\left(|1\rangle+|6\rangle\right)$ (the superposition
of excited sites $1$ and $6$).

\begin{figure}
\begin{centering}
\includegraphics[width=1\textwidth]{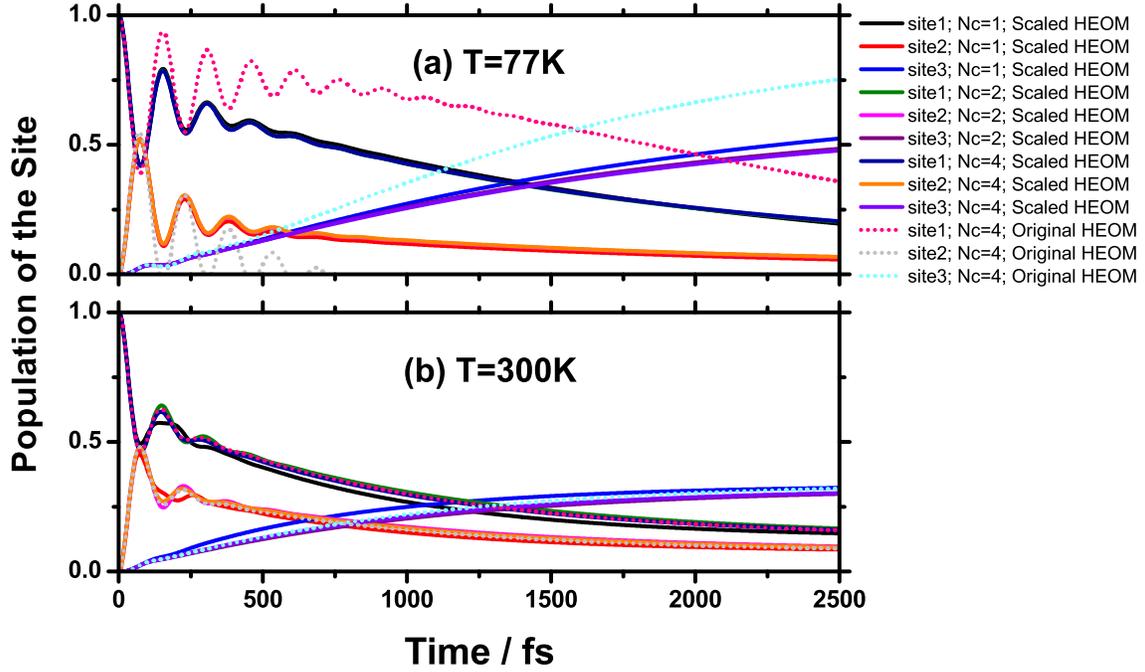} 
\par\end{centering}

\caption{The population evolution of site $1$, $2$ and $3$ under different
cutoffs for the tier of auxiliary density operator ($\mathcal{N}_{c}$)
and different HEOM approaches. The solid lines represent the population
evolution at three different truncation levels $\mathcal{N}_{c}=1$,
$2$ and $4$ of the scaled HEOM approach. The short dot lines show
the time evolution of $\mathcal{N}_{c}=4$ at the original HEOM approach.
Site $1$ is initially excited and the reorganization energy and Drude
decay constant are $\lambda_{j}=\lambda=35\;\unit{cm^{-1}}$ and $\gamma_{j}^{-1}=\gamma^{-1}=50\;\unit{fs}$,
respectively. The dynamics are shown at cryogenic temperature $T=77\unit{K}$
(upper panel) and at physiological temperature $T=300\unit{K}$ (lower
panel). \label{cutoffTest} }

\end{figure}

\textit{Calibration}$-$ We compare the scaled approach to the original
HEOM approach and investigate the critical choice of the truncation
levels of ADOs ($\mathcal{N}_{c}$) and the correlation functions
($K$). The original HEOM approach is given in Eq. (\ref{eq: HEOM}).
In Fig. \ref{cutoffTest}, we depict the population of sites $1$,
$2$ and $3$ for different $N_{c}$ in the scaled approach and for
$N_{c}=4$ in the original approach at temperatures $T=77\;\unit{K}$
and $T=300\;\unit{K}$, setting $K=0$. One obtains a large difference
between the two approaches at $T=77\;\unit{K}$, see the dotted lines
in the figure. Under the original HEOM method, the population of site
$2$ goes below $0$ after $750\;\unit{fs}$, which is unphysical.
However, the population of each site under the scaled HEOM approach
behaves reasonably even at $\mathcal{N}_{c}=1$. This shows that the
scaled HEOM approach can result in better simulation results at less
computational costs. The difference between the two approaches originates
from the truncation level of the correlation function, which is due
to the coupling between the system and the environment. 

For the scaled approach itself, the difference between different truncation
levels ($\mathcal{N}_{c}$) is modest at both temperatures. It can
be seen that there is only minimal difference among three $\mathcal{N}_{c}$
values at $T=77\;\unit{K}$. Beyond $1500\unit{\; fs}$, the difference
of the population evolution for site $3$ becomes larger between the
case of $\mathcal{N}_{c}=1$ and $\mathcal{N}_{c}=2$ and $4$. The
population evolution of all the sites is exactly the same for $\mathcal{N}_{c}=2$
and $4$. As a result, $\mathcal{N}_{c}\geq2$ is the sufficient truncation
level for the ADOs at $T=77\;\unit{K}$. At room temperature, $T=300\unit{\; K}$,
the variation at the dynamics of the populations as a function of
$\mathcal{N}_{c}$ is more apparent. The population evolution of all
sites for $\mathcal{N}_{c}=1$ is not as smooth as in the other two
situations. For the cases of $\mathcal{N}_{c}=2$ and $4$, there
is a slight difference in the population beatings which occur between
$200\;\unit{fs}$ and $300\;\unit{fs}$. Although this is not a substantial
difference, we believe $\mathcal{N}_{c}=4$ is good compromise between
efficiency and accuracy.

For the truncation level of the correlation function ($K$), the simulation
for the seven sites is computationally unwieldy. Therefore, we truncated
the system to test the correlation truncation level using a three-site
model (sites $1$, $2$ and $3$) with three different values of $K$,
being $K=0,\;1\;\text{{and}}\;2$. The results show that truncation
level $K=0$ is enough for both $T=77\;\unit{K}$ and $300\;\unit{K}$.
A similar result was also found in \cite{Shi2009a}, where non-zero
K was shown to be significant in the dynamics only at rather long
time scales. In the following computations, we choose $\mathcal{N}_{c}=4$
for both temperatures as our reference. At this point, we would like
to emphasize the numerical efficiency of the scaled HEOM approach.
The original HEOM approach requires a truncation level as high as
$\mathcal{N}_{c}=12$ to get converged results \cite{Fleming-PNAS-2009}.
However, we only require $\mathcal{N}_{c}=2$ for $T=77\;\unit{K}$
and $\mathcal{N}_{c}=4$ for $T=300\;\unit{K}$, which is a significant
resource reduction. On a standard desktop computer, a simulation of
the time-evolution for $2.5$ ps takes about $7$ minutes for the
case of $\mathcal{N}_{c}=2$ and about $1.5$ hours for the case of
$\mathcal{N}_{c}=4$.

\begin{figure}
\begin{centering}
\includegraphics[width=0.8\textwidth]{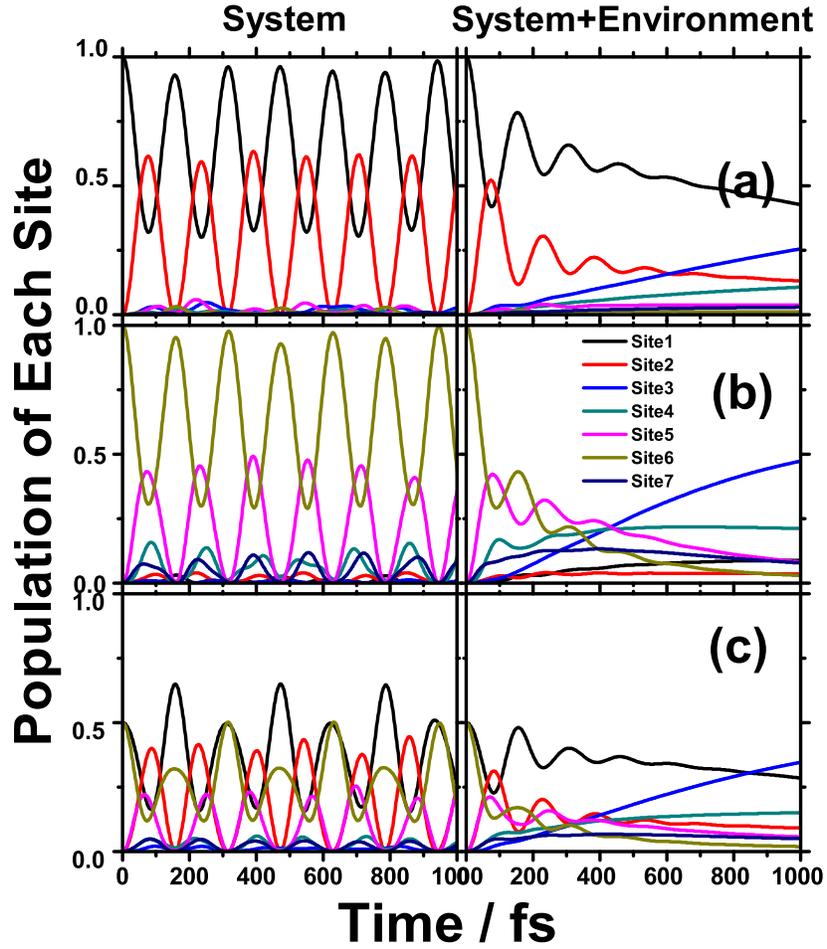} 
\par\end{centering}

\caption{The population evolution of each site at cryogenic temperature, $T=77\;\unit{K}$.
The left panel shows the dynamics for the system alone and the right
includes the effects of the environment. The reorganization energy
is $\lambda_{j}=\lambda=35\;\unit{cm^{-1}}$, while the value of Drude
decay constant is $\gamma_{j}^{-1}=\gamma^{-1}=50\;\unit{fs}$. The
initial conditions are site $1$ excited $\left(a\right)$, Site 6
excited $\left(b\right)$ and the superposition of site 1 \& 6 $\left(c\right)$.
\label{77K}}

\end{figure}

\textit{Coherent beatings at cryogenic and room temperatures}$-$
Now, we investigate the cryogenic temperature $T=77\;\unit{K}$ in
more detail. This being the temperature of the first experiment by
Engel et al. \cite{Engel-2007} which shows coherent phase evolution
of the FMO complex from time $t=0$ to roughly $t=660\;\unit{fs}$.
The results are presented in Fig. \ref{77K}. On the left panel, we
show the results of simulation for the system Hamiltonian only. The
the right panel one observes that the quantum beating between certain
sites clearly persists in the short time dynamics of the full FMO
complex. For the simulated initial conditions, the population beatings
can last for hundreds of femtoseconds; this time scale is in agreement
with the experimental observation \cite{Engel-2007}. The population
beating for all three different initial conditions can last around
$650\;\unit{fs}$. In Fig. \ref{77K}(a), the initial state is localized
at site $1$. The system exhibits coherent beatings between the strongly
coupled sites 1 and 2, accompanied by relatively slow relaxation to
sites $3$ and $4$. The change of population of all other sites is
weak. In Fig. \ref{77K}(b), where the initial state is localized
at site $6$, the population relaxes faster. For $t\leq400\unit{\; fs}$,
there is population beating between the strongly coupled sites $6$
and $5$, accompanied by relaxation to the intermediate sites $4$,
$5$ and $7$. From these sites, the population is fed into the low-energy
sites $3$ and $4$. The population of site $6$ almost vanishes at
$t=800\;\unit{fs}$, while for the previous initial condition, Fig.
\ref{77K}(a), the population of site $1$ is roughly $0.5$ at that
time. The exciton migration pathways and time scales are in accordance
with previous work \cite{Adolphs2006,Fleming-PNAS-2009,Fleming-Science-2007}.
Finally, Fig. \ref{77K}(c) represents the superposition of site $1$
and $6$. The time evolution of this case is the combination of the
single site excited cases. That is the population evolution on site
1 follows the pathway of initially single site 1 excited, while the
pathways for the population on site 6 is the same as the single site
6 excited case.

\begin{figure}
\begin{centering}
\includegraphics[width=0.8\textwidth]{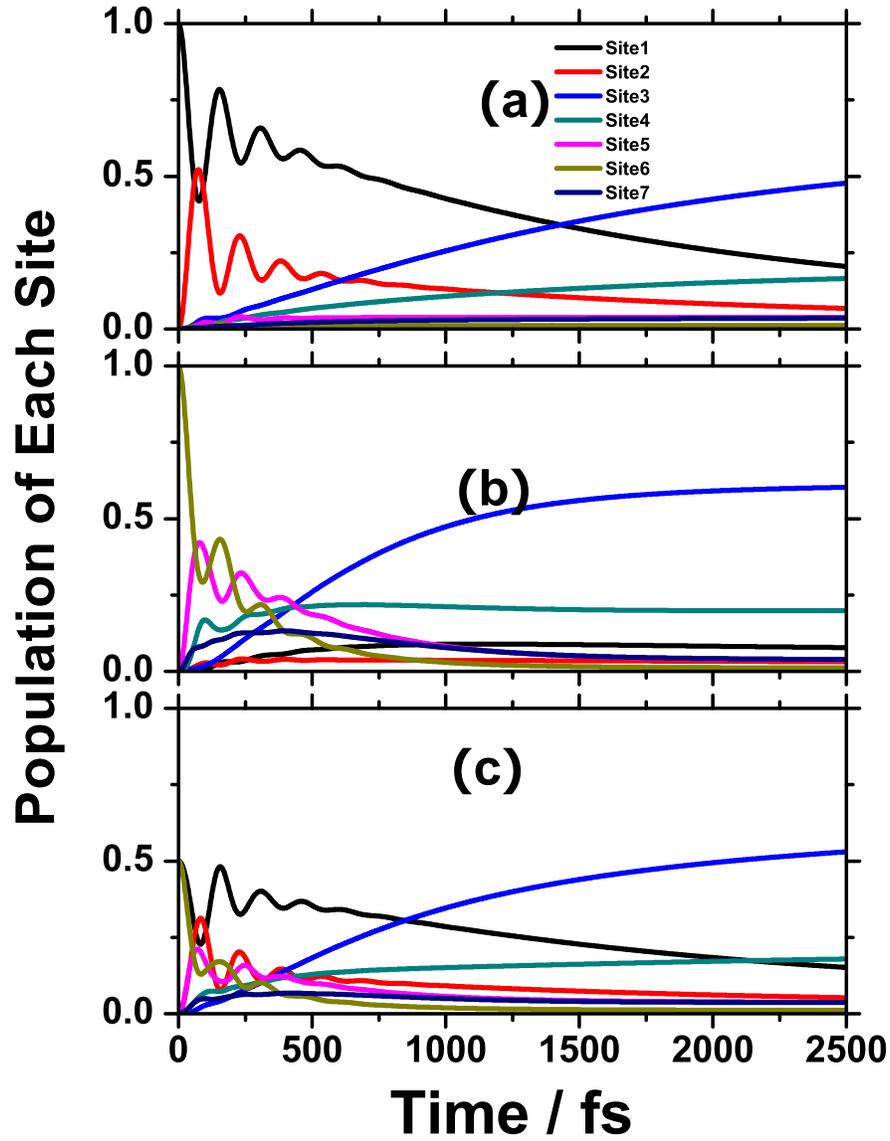} 
\par\end{centering}

\caption{Long time-dynamics of the population at each site for $T=77\;\unit{K}$,
where $\left(a\right),\;\left(b\right)$ and $\left(c\right)$ are
corresponding to different initial conditions as noted before. All
other parameters are the same as Fig. \ref{77K}. \label{longtime}}

\end{figure}

In order to investigate the excitation transfer beyond the initial
beating region, we extended the simulation to a longer time $\left(\sim2500\;\unit{fs}\right)$.
The result are shown in Fig. \ref{longtime}. To check whether the
entire system converges to the thermal equilibrium, we obtained all
eigenvalues of the system Hamiltonian and calculate the probability
of each eigenstates under temperature $T$ based on the Boltzmann
distribution. Subsequently, we transformed the population from the
eigenstate representation to site representation and obtained the
population of each site at thermal equilibrium. At $T=77\unit{K}$,
the population of site $3$ is $0.69$ and that of site $4$ is $0.22$.
The population of all other sites is smaller than $0.03$. For the
case of having the initial excitation start in site $6$, the thermal
equilibrium is reached the end of $2\unit{ps}$, while for the case
of the other two initial conditions, they are still on their way to
the thermal equilibrium at the end of $2.5\unit{ps}$. Our simulation
shows that the system reaches thermal equilibrium at $\sim7\;\unit{ps}$
for the case in which site $1$ was initially excited and the time
for the initial superposition of site $1$ and $6$ is around $6\;\unit{ps}$.

\begin{figure}
\begin{centering}
\includegraphics[width=1\textwidth]{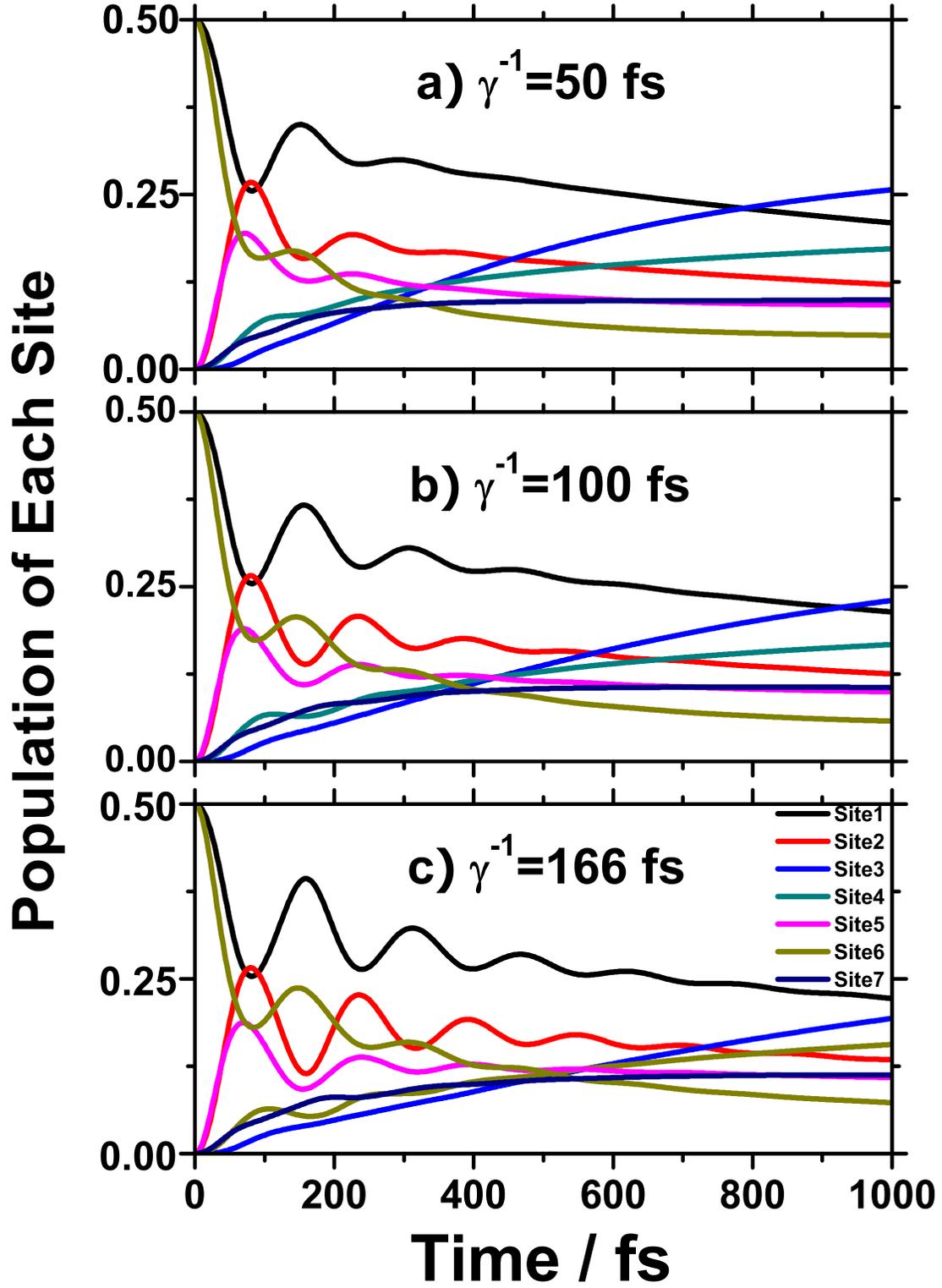} 
\par\end{centering}

\caption{The population of all FMO sites at $T=300\;\unit{K}$. The initial
state is the superposition of site $1$ and $6$. The reorganization
energy remains $35\;\unit{cm^{-1}}$. Three different values of phonon
relaxation time are tested, which are $\gamma^{-1}=50\;\unit{fs}\,\left(a\right)$
, $\gamma^{-1}=100\;\unit{fs}\,\left(b\right)$ and $\gamma^{-1}=166\;\unit{fs}\,\left(c\right)$.
\label{300Kgamma}}

\end{figure}

Recent experiment studied the excitation dynamics of the FMO complex
at room temperature \cite{Engel2}. To investigate quantum coherence
effects under physiological conditions, we simulate the dynamics at
the temperature $T=300\;\unit{K}$. We choose three different values
for $\gamma^{-1}$ in our calculation: $50\;\unit{fs},\,100\;\unit{fs}$,
and $166\;\unit{fs}$ \cite{Fleming-JPC-2005,Fleming-PNAS-2009}.
Following the same procedure as before, we consider three different
initial conditions with the reorganization energy $\lambda=35$ cm$^{-1}$.
We choose the truncation level at $\mathcal{N}_{c}=4$. The calculation
results are shown in Fig. \ref{300Kgamma}.

The main difference between the case of $300\;\unit{K}$ and that
of $77\;\unit{K}$ is the time scale of the persistence of population
beating. The coherent beating lasts only $400\;\unit{fs}$ at room
temperature whereas it lasts much longer at $T=77\;\unit{K}$. It
is also found that the smaller $\gamma$ is, the longer the population
beating can last. When $\gamma^{-1}=166\;\unit{fs}$, the population
beating time can last almost $700\;\unit{fs}$. The main pathways
for all cases are the same as the pathways at low temperature. Furthermore,
the time evolution of the population for each site also converges
to thermal equilibrium. However, the entire system reaches thermal
equilibrium considerably sooner at room temperature. For example,
the system initialized at site $6$ reaches equilibrium at $1.5\;\unit{ps}$
when $T=300\;\unit{K}$, compared to $2\;\unit{ps}$ at $77\;\unit{K}$.

\begin{figure}
\begin{centering}
\includegraphics[width=1\textwidth]{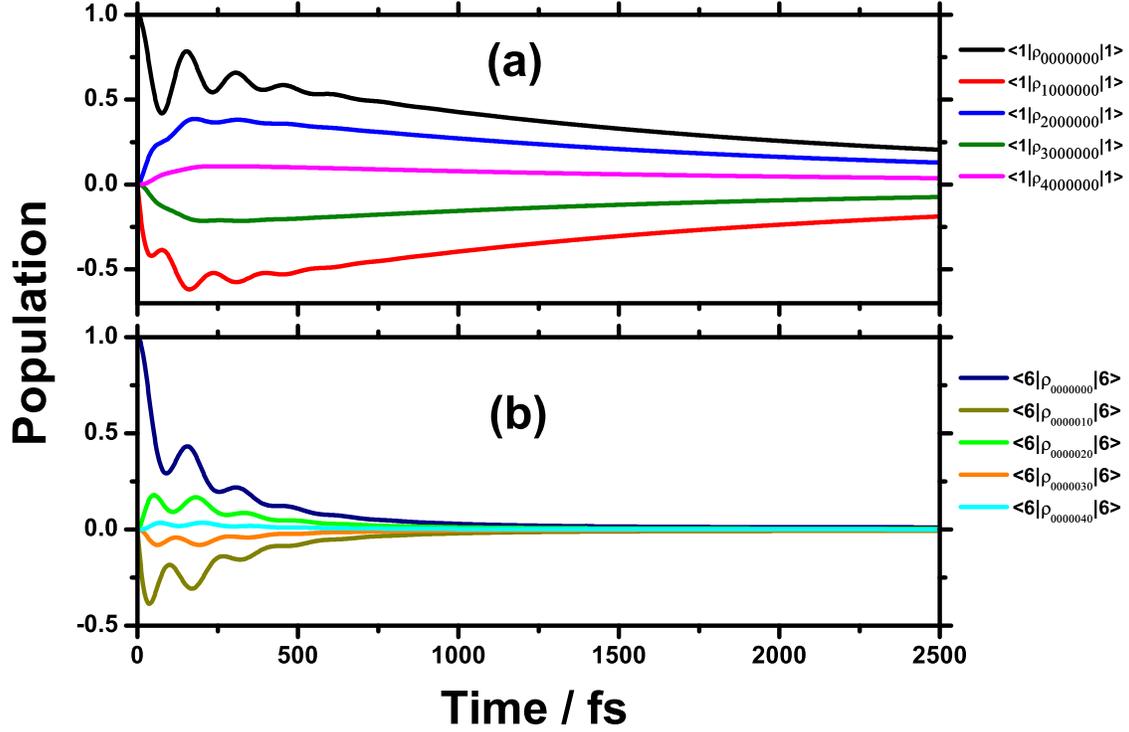} 
\par\end{centering}

\caption{The population evolution of RDO and ADOs for the case of initial excitation
at sites $1$ and $6$ respectively. The first panel shows the time
evolution of the ADO $\langle1|\rho_{n000000}|1\rangle$ elements
with $n=0,\;1,\;2,\;3,\;\mbox{and }4$. The second panel shows the
time evolution of the $\langle6|\rho_{00000m0}|6\rangle$ elements
with levels of truncation $m=0,\;1,\;2,\;3,\;\mbox{and }4$.\label{ADOs}}

\end{figure}

\textit{Behavior of the auxiliary density operators}$-$ In order
to further investigate the effects of the ADOs and their role in the
modified HEOM scheme, we examined the magnitude of their population
elements and found that the majority of them are close to $0$. However,
there are some non-zero ADO elements during the time evolution. We
plot their time dependence in Fig. \ref{ADOs}. For the case where
site $1$ is initially excited, the simulation shows that the most
important ADO elements are $\langle1|\rho_{n000000}|1\rangle$ with
$n=1,\;2,\;3,\;\mbox{and }4$. While for the site $6$ initially excited
case, the important ADOs are $\langle6|\rho_{00000m0}|6\rangle$ with
$m=1,\;2,\;3,\;\mbox{and }4$. From the image (Fig. \ref{ADOs}),
it can be found that the amplitude of the ADOs decays rather quickly
as the level of truncation increases. Conversely, the ADO populations
are related to the amplitude of the site population in RDO. When the
population goes up, the corresponding population in ADOs also increases.
For odd truncation levels ($\rho_{1000000}$, $\rho_{3000000}$, $\rho_{0000010}$
and $\rho_{0000030}$), the ADOs yield negative population. The results
are indicative of the fact that the scaled HEOM approach reduces the
amplitude of ADOs as truncation level increases. Interestingly, there
are no negative population elements of the density matrix when the
truncation level of the correlation function is $K=0$ at cryogenic
temperature. This is in contrast to the original hierarchical approach
\cite{Fleming-PNAS-2009}, in which some populations become negative. 

\begin{figure}
\begin{centering}
\includegraphics[width=0.8\textwidth]{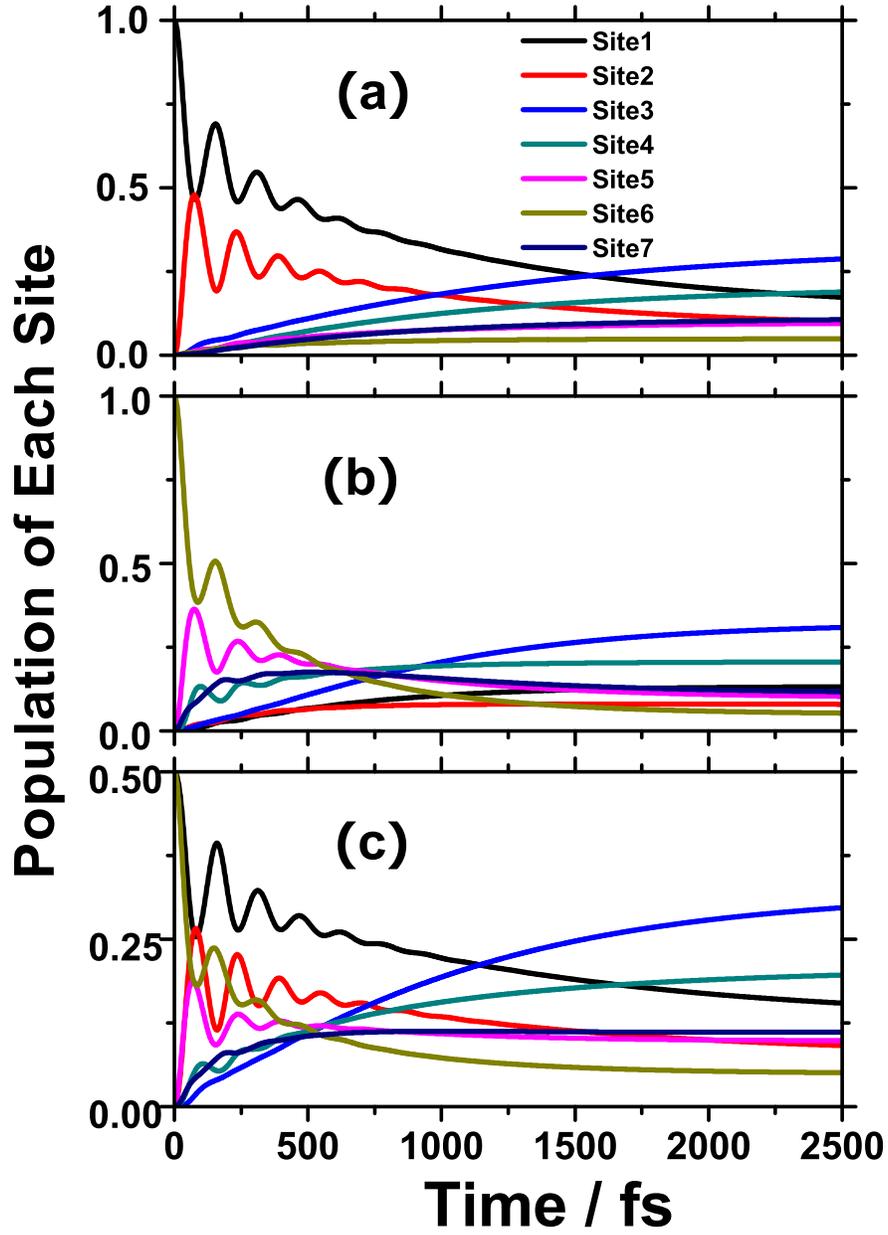} 
\par\end{centering}

\caption{Long time-evolution of the population of each site at $T=300\;\unit{K}$,
where $\left(a\right),\;\left(b\right)$ and $\left(c\right)$ correspond
to site $1$ initially excited, site $6$ excited and the superposition
of site $1$ and $6$. The reorganization energy is $\lambda_{j}=\lambda=35\;\unit{cm^{-1}}$,
and $\gamma_{j}^{-1}=\gamma^{-1}=166\;\unit{fs}$. \label{longtime300K}}

\end{figure}

\textit{Energy transfer pathways}$-$ We briefly comment on the exciton
transfer pathways. The pathways are determined by the system Hamiltonian
rather than by the system-environment coupling or the environment
\cite{Adolphs2006,Fleming-PNAS-2009}. From Fig. {[}\ref{77K}, \ref{longtime},
\ref{300Kgamma} and \ref{longtime300K}{]}, we can find that the
frequency of site population oscillation is independent of the temperature
and different bath relaxation. For example, in the pathway $\mbox{site }1\rightleftharpoons2\longrightarrow3\&4$
the main beatings between site $1$ and $2$ are caused by several
features. The energy barrier between site $1\&2$ ($\Delta\varepsilon_{12}=-120\;\unit{cm^{-1}}$)
is smaller than that of site $1\&6$ ($\Delta\varepsilon_{16}=-220\;\unit{cm^{-1}}$)
and the coupling of sites $1\&2$ is stronger. The population oscillation
between site $3\&4$ is due to the similar site energies and the strong
coupling between them. In the real biological system, the RC is close
to site $3$ and $4$. When the exciton transfers to these sites,
it moves to the RC directly and cannot return to the system. Other
pathways start from site $6$, i.e. $\mbox{site }6\rightleftharpoons7\longrightarrow3\&4$,
$\mbox{site }6\rightleftharpoons5\longrightarrow3\&4$ and $\mbox{site }6\longrightarrow3\&4$.
Although the population distribution is the same for all pathways
at long times, the excitation transfer time is shorter for an initial
excitation at site $6$ \cite{Adolphs2006}. For the pathway of site
$1$, the site energy of site $2$ is bigger than that of site $1$.
It is hard for the excitation to move from site $1$ to site $2$
and that is why the wave-like evolution lasts for a longer time. However,
in the population pathway that starts from site $6$, the excitation
flows from the higher- to lower- energy sites all the time. This reduces
the transfer time and the system reaches thermal equilibrium faster.
Comparing the two different temperatures in terms of the excitation
transfer pathways, we note that the influence of the thermal bath
is much stronger at room temperature than at low temperature. The
bath at $300\;\unit{K}$ helps the system to transfer the excitation
more efficiently by reducing the quantum beating, thus speeding up
the overall transfer times to the reaction center.

\section{Conclusion}

In summary, we have examined the full dynamics of the transfer of
excitation energy towards the reaction center through the Fenna-Matthews-Olson
(FMO) pigment-protein complex, employing the modified scaled hierarchical
approach recently developed by Shi et al. \cite{Shi2009a}. The scaled
HEOM approach not only reduces the cutoff for the tier of auxiliary
density operators, but also decreases the truncation level of the
correlation function, which makes it more efficient compared to the
original HEOM approach. We have shown that a tier cutoff of $N_{c}=4$
and a correlation function cutoff of $K=0$ optimizes simulation efficiency
and accuracy for the parameter regime of the FMO complex. Furthermore,
our theoretical results show that the population beating can last
as long as $650\;\unit{fs}$ under cryogenic temperature ($77\;\unit{K}$).
When the temperature is $300\;\unit{K}$, the beating time can vary
from $400\;\unit{fs}$ to $700\;\unit{fs}$, depending on the environment
parameters. Our simulation result is in accord with the conclusion
of Ishizaki et al \cite{Fleming-PNAS-2009}. The improved computational
performance of our scaled HEOM approach will be especially useful
in theoretical studies of transport measures such as: efficiency;
transfer time; and other properties, such as entanglement. Moreover,
this efficient approach also provides us with the potential to couple
other effects into our current system. Under the current model, only
the thermal effect is fully considered; however, there exist many
other effects in the real biological system such as: dipole-dipole
interaction; the different phonon environment for each site; and slow
structure changes of the FMO complex. It will be our future task to
build a model with these features and examine the time evolution of
entanglement and related quantum information measures \cite{Kais2007,Wei2010,Xu2010}.

\section{Acknowledgment}

We thank the NSF Center for Quantum Information and Computation for
Chemistry, Award number CHE-1037992. J.Z. and S.K. acknowledge the
ARO for financial support. A.A.-G. and P.R. were supported as a part
of the Center for Excitonics, as an Energy Frontier Research Center
funded by the U.S. Department of Energy, Office of Science, Office
of Basic Energy Sciences under Award number DE-SC0001088. A.A.-G.
acknowledges the award entitled {}``Coherent Quantum Sensors\textquotedbl{}
through the Defense Advanced Research Projects Agency program, Quantum
Effects in Biological Environments.

\bibliographystyle{achemso} \bibliographystyle{achemso}
\bibliography{Scaled_Hierarchy_Approach}

\newpage{} %%%%%%%%%%%%%%%%%%%%%%%%%%%%%%%%%%

%%%%%%%%%%%%%%%%%%%

%

%%%%%%%%%%%%%%%%%%%%%%%%%%%%%%%%%%

\end{document}